\documentclass[aps,pre,twocolumn,superscriptaddress,amssymb,showpacs,showkeys]{revtex4}
\usepackage{graphicx}
\def\d{\partial}
\def\be{\begin{eqnarray}}
\def\ee{\end{eqnarray}}

\begin{document}
\title{Are Scale Free Networks Better?}
\author{Gergely P\'eli}
\email{ply@cs.elte.hu}
\affiliation{Communication Network Laboratory, E\"otv\"os University,\\
P\'azm\'any P. 1/A, Budapest 1117, Hungary}
\author{G\'abor Papp}
\email{pg@ludens.elte.hu}
\affiliation{Communication Network Laboratory, E\"otv\"os University,\\
P\'azm\'any P. 1/A, Budapest 1117, Hungary}
\affiliation{Department for Theoretical Physics, E\"otv\"os University,\\
P\'azm\'any P. 1/A, Budapest 1117, Hungary}

\date{\today}

\begin{abstract}
We study the performance of Weibull and scale free Internet-like
networks and compare them to a classical random graph based network. 
The scaling
of the traffic load with the nodal degree is established, and confimed
in a numerical simulation of the TCP traffic. The scaling allows us to
estimate the link capacity upgrade required making and extra
connection to an existing node.
\end{abstract}

\maketitle

\section{Introduction}

Random graphs were studied since the middle of the 20th century with
the initiative works of mathematicians P\'al Erd\H os and Alfr\'ed
R\'enyi. In 1960 they published their paper
{\it On the evolution of random graphs}, 
with the first thorough study of the graph theory~\cite{er} . However, these
results were strictly theoretical, since tools to measure real-world
random graphs at that time were not available. With the evolution of
the personal computers, we have now the possibility to study this
topic in practice, and, interestingly, the computers themselves
provide one of the most exciting real random graphs: the 
computer networks. 

As these networks evolved, their properties were analyzed, and soon it
was clear, that the Erd\H os-R\'enyi models (ER) were not appropriate for 
these graphs~\cite{www}. The most important difference was the
\emph{scale free} 
property, that the computer networks have, but the classical models
lack. It turned out, that this feature is shared by other types of
networks, like social~\cite{social} and metabolic~\cite{meta} ones.
It means that the distribution of the degrees of the nodes
follows a power-law distribution, while the classic graphs has a
Poisson degree distribution, with an exponential tail. It shows that,
in the real-world graphs, nodes with a high number of connections are
much more likely than expected. Therefore new models were necessary to
describe these kind of random graphs.

One of these models was developed by L\'aszl\'o Barab\'asi and R\'eka
Albert (BA model)~\cite{alb,abr}. Their method has two key features:
\emph{incremental growth} and 
\emph{preferential attachment}. Incremental growth means that the
graph is constructed by adding nodes to the existing graph, and
connecting them according to a construction rule, contrary to the
original Erd\H os-R\'enyi picture with a static graph.
Preferential attachment means the the likelihood of a connection
depends on the degree of a node, again lifting the classical
assumption of equal probabilities. This model has more variations, and
they are able to describe a wide class of random graphs. There is
another class of models study such networks without incremental
growing, purely on their statistical properties~\cite{zck}.

In this paper, we investigated network properties of BA-like models
with different parameters and raise the question whether non-classical
graphs may perform better transmitting data over them. Certain value of
parameters yield a graph with an 
exponential tail degree distribution, hence allowing us to compare
these classical type models to the ones with power-law distribution. 
In Section II we analyze the properties of the extended BA models,
estimating their node distribution. Next, in Section III we construct
networks based on these models, and simulate a network traffic on such
a graph in a simplified model estimating the traffic load on the
nodes. The scaling of the load with nodal degree is presented,
allowing to estimate the proper bandwidth allocation when upgrading a
node. In Section IV we study a more realistic setup with TCP dynamics
and compare the theoretical result of the previous section to the
simulated ones, while in Section V we discuss the overall performance
of the different simulations. Finally, we conclude out
analysis.

\section{The model}

In the original BA-model the graph is constructed as follows. Starting
from a small initial graph, we extend it by adding a new node in
each step and connecting it to $m$ randomly selected existing
nodes. The probability of choosing a particular node is proportional
to its degree,
\be
  p_i=\frac{d_i}{\sum_{j=1}^{n-1} d_i}\,.
\ee
This model constructs a scale free graph, where the cumulative degree
distribution has a power-law decay with an exponent 2~\cite{alb}.

In the following, we modify the construction law, and use the more
general form
\be
  p_i=\frac{d_i^\alpha}{\sum_{j=1}^{n-1} d_i^\alpha} \,,
\ee
weighting the probability with a power $\alpha$ of the nodal
degree, similarly to Ref.~\cite{krapivsky}. The BA model 
corresponds to $\alpha$=1, while for $\alpha=0$ the preferential
connectivity is cancelled, and we are back to a classical ER-like
graph model with a uniform distribution, leading to exponential node
distribution. While this model is similar to the original ER model, it
differs in some aspects, such as it has a minimal guaranteed degree,
and the ordering of the nodes presents nonzero correlations in the
degrees~\cite{callaway}. With $\alpha\in(0, 1)$, the models provide a
smooth transition between the classical and the scale free models.

The degree distribution for these models can be derived following
the method introduced in~\cite{alb}. Here we give a fast estimate
on the distribution, for the exact result see~\cite{krapivsky}. 
First, we estimate the rate of
growth of the degree at each node, assuming that the growth of the degree is
continuous in time. At time $t$ there are exactly $t$ nodes and $m\,t$
links between them. Hence, the expected degree value $k_i=E(d_i)$ 
of node $i$ is growing as
\be
  \d_t k_i(t) = \frac{m\,k_i(t)^\alpha}{\sum_{j=0}^t k_j(t)^\alpha}
  \,.
\label{ba-ext}
\ee
For $\alpha=0$, the denominator is simply conunting the number of
nodes, and is equal to $t$, while for $\alpha=1$, it is (double)
counting the number of links, and thus is $2m\,t$, both being a linear
function of the time $t$. Numerical simulations showed that for 
$\alpha\in(0, 1)$ the denominator is well approximated by a linear
function $c\,t$, where $c$ is the $\alpha$ dependent measured 
slope. Hence, generally our differential equation can be written as
\be
\d_t k_i(t) = \frac{k_i(t)^\alpha}{c\,t}\,.
\label{weib-appr}
\ee
Fortunately, these equations are solvable for each $\alpha$ in the
choosen $(0,1)$ range. Specifically, for $\alpha=0$ we get
\be
  k_i(t)=\frac{\log t}{c} \,,
\ee
and more generally for $\alpha\in(0,1)$
\be
  k_i(t)=\left((1-\alpha) \frac{\log t}{c}\right)^{\frac{1}{1-\alpha}}\,.
\label{kalpha}
\ee
Finally, for the scale free case $\alpha=1$ the structure changes, and
we arrive at
\be
  k_i(t)=t^{\frac1c}\,.
\label{kpower}
\ee
In order to get a formula for the degree distribution, note that for
each $i$, $k_i(i)=m$, since at the time the node is added it has
exactly $m$ connections. The growth process for node $i$ has a
similarity to the growth of the previous nodes, and this suggests that
\be
  k_i(t) = k_1\left(\frac{t}{i}\right)\,.
\ee
The cumulative distribution $P(k_i >x)$ counts the number of nodes
with $k_i(t)>x$, i.e. $k_1\left(\frac{t}{i}\right) > x$, which leads
to
\be
  \frac{t}{k_1^{-1}(x)} > i\,.
\ee
Normalizing back with the total number of nodes at time $t$ (being
also $t$) we get the approximate cumulative probability
\be
  P(k_i > x) \simeq \frac1{k_1^{-1}(x)} \,.
\ee
Inverting Eq.~(\ref{kalpha}) we find
\be
  P(k_i > x) \simeq e^{-\frac{c}{1-\alpha}\,x^{1-\alpha}}\,,
\label{weib}
\ee
the Weibull-distribution for $\alpha\in(0,1)$ with scale parameter $c$,
and shape parameter $1-\alpha$.  We note, that the exact result 
is of the form~\cite{krapivsky} 
\be
   P(k_i > x) \simeq x^{-\alpha} e^{-\frac{c}{1-\alpha}\,x^{1-\alpha} ...}\,,
\ee
where $...$ denotes higher order correction terms.

For $\alpha=0$ ($c=1$) one recovers
the classical limit $P(x)\sim e^{-x}$. In the region $\alpha\in(0,1)$ the
distribution still vanishes exponentially, so strictly speaking no
heavy tails are present, however, the probability of finding nodes
with large degree increases dramatically. Numerical studies revealed,
that the distribution indeed follows the form~(\ref{weib}), however,
the parameters are slightly different from the predicted ones, due to the
approximation used in~(\ref{weib-appr}). In the limit $\alpha=1$
($c=2$) we arrive at the genuine scale free heavy tail result
\be
  P(x) \sim x^{-c} \,.
\ee
We note, that the analysis can be extended to $\alpha>1$, however,
such a distribution actually tends to the degenerate case with one
node being connected with all others in a star-like topology. For
$\alpha=1$ the original BA model results in $c=2$, however, different
growth stategies are able to vary the value of $c$, hence the decay
exponent. In this work we restrict ourselves to the original BA model.

\section{Estimation of the network load in a simplified model}

\begin{figure*}
\includegraphics[width=0.32\textwidth]{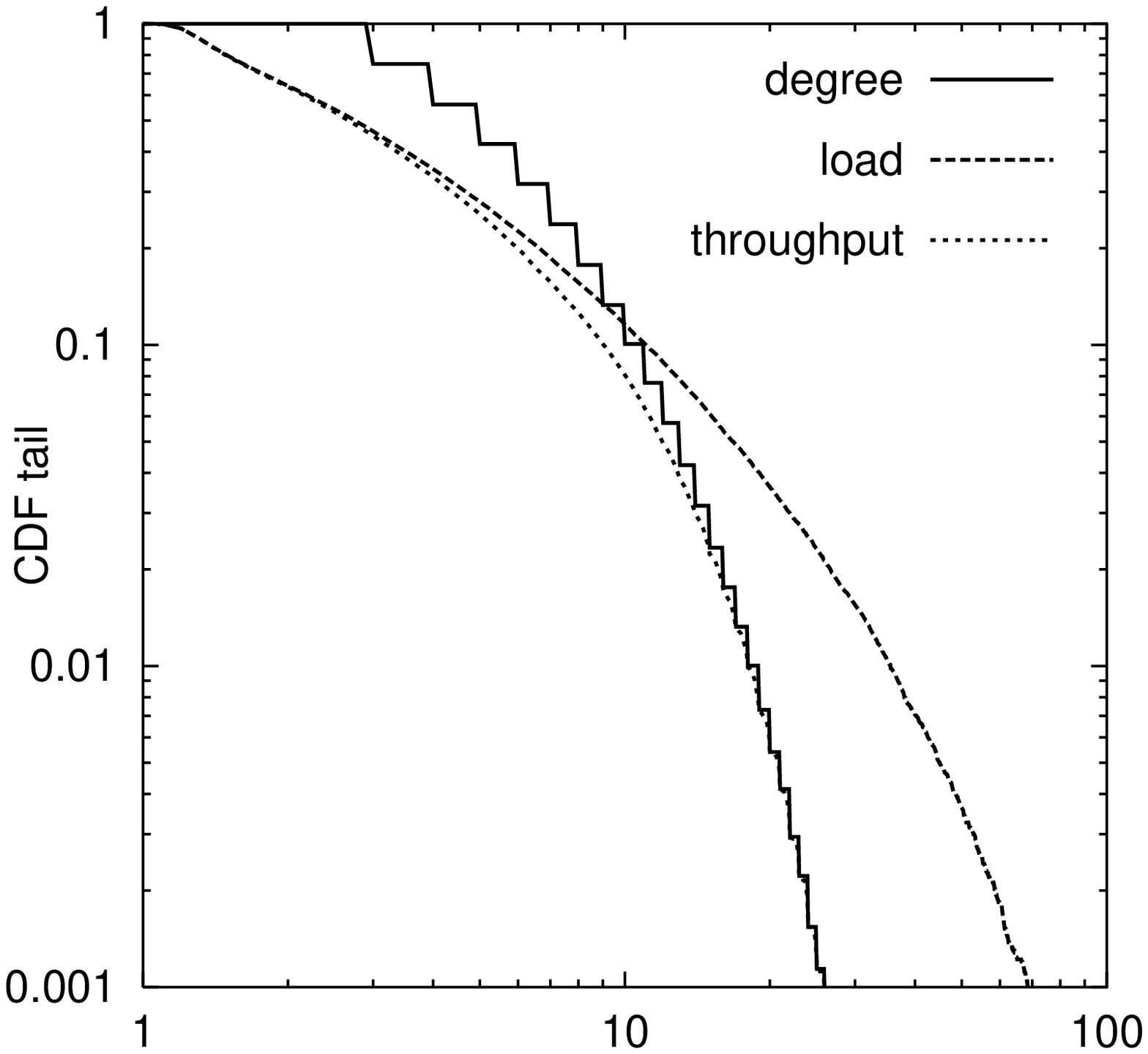}
\includegraphics[width=0.32\textwidth]{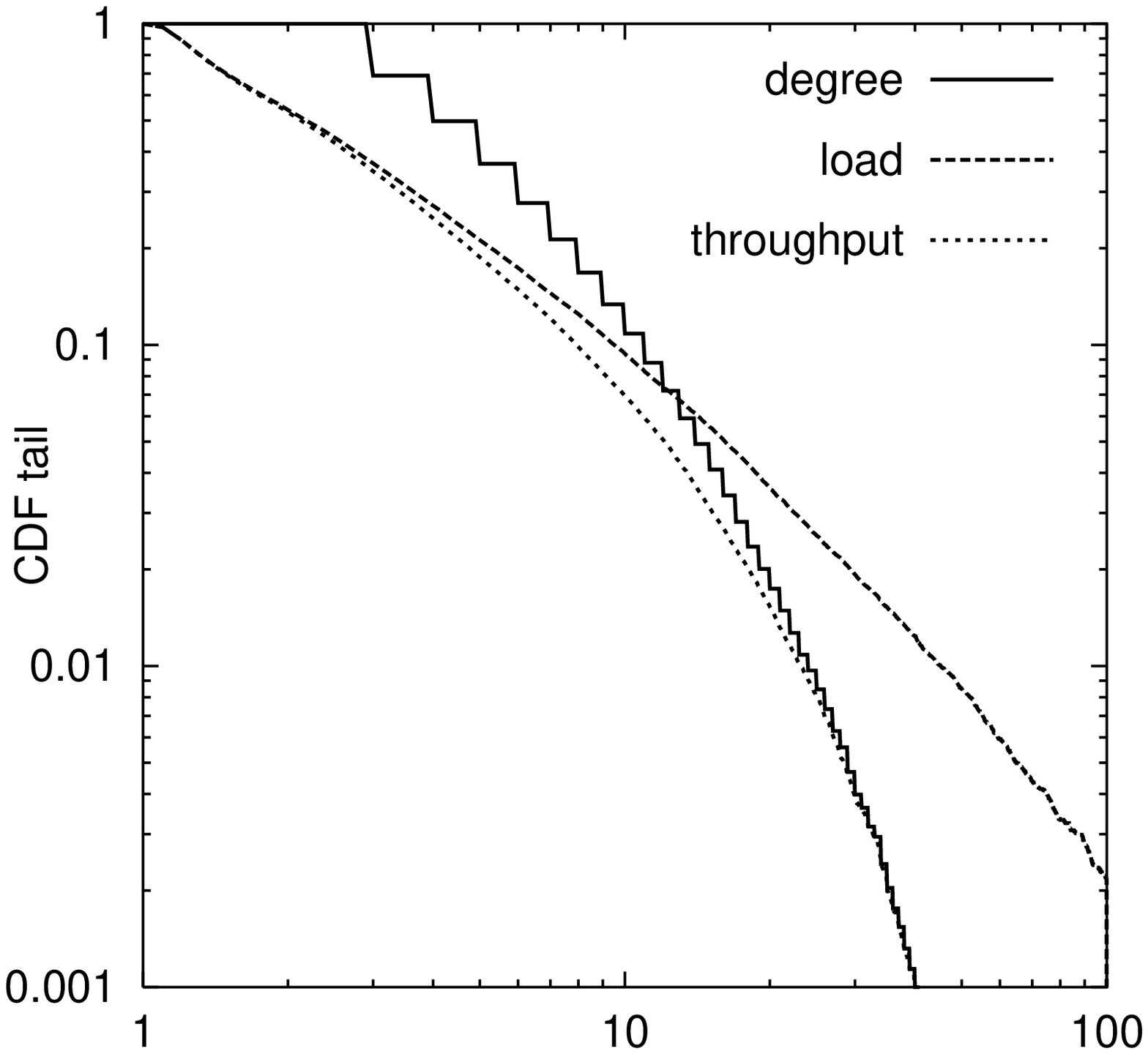}
\includegraphics[width=0.32\textwidth]{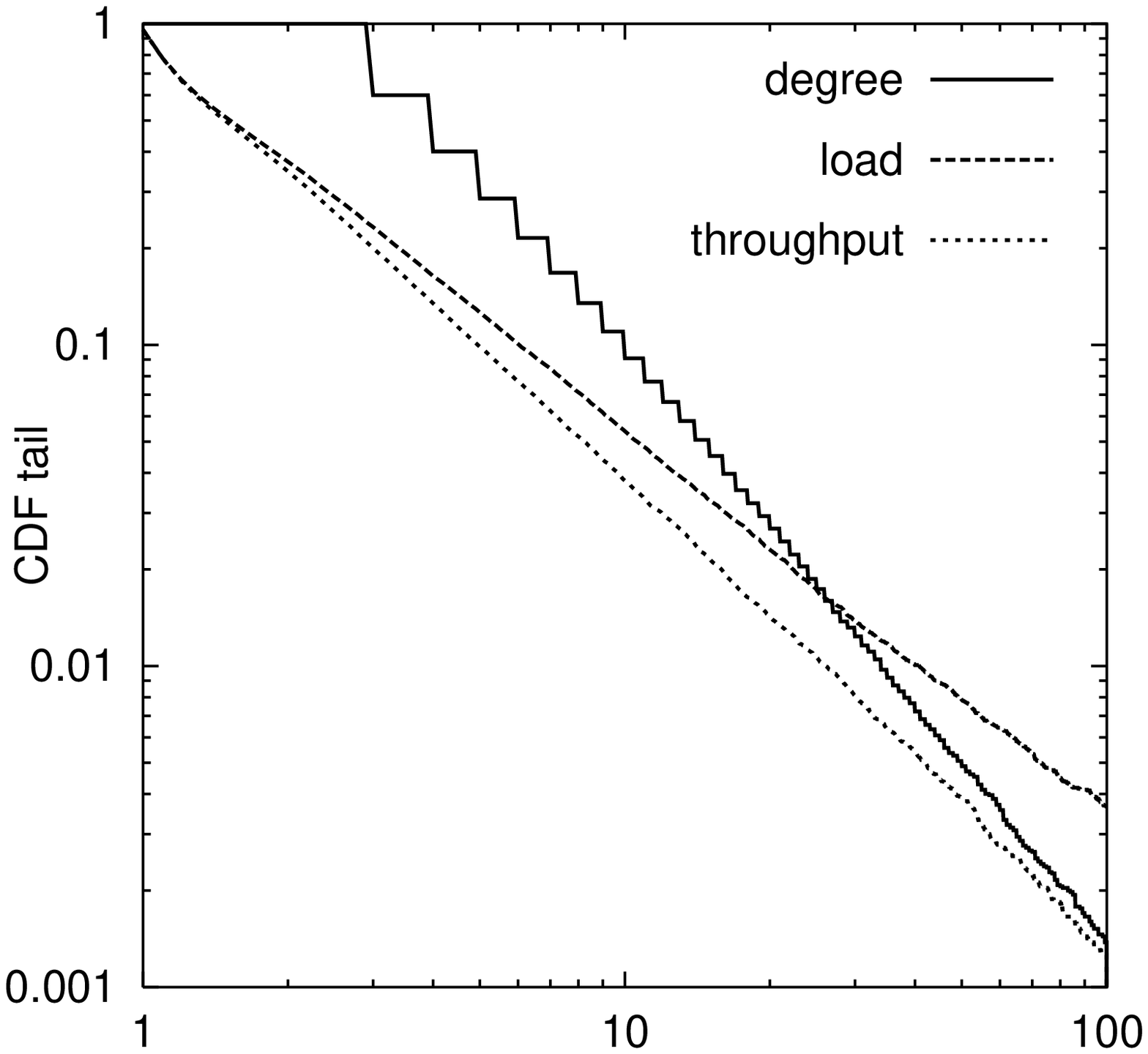}
\caption{Tail of cumulative distributions of the nodal degree, load
  and throughput for geometries with $\alpha=$0 (ER graph, left), 
0.5 (middle) and 1 (BA graph, right).}
\end{figure*}

In the following, first we generate random graphs on a computer
according to the extended BA model~(\ref{ba-ext}) and study the load
in a simplified network model. We assume, that each node generates a
constant traffic (data stream) to all the other nodes. The
amount of 
traffic is the same for all connections, and furthermore, we assume
that the link capacity can handle the accumulated traffic. The
properties of the network are analysed tuning the geometrical
parameter $\alpha$.

For each value of $\alpha$ we performed measurements on 8 different
graphs (generated with the same statistical properties), and then
averaged. Each topology was made out of $N=100000$ nodes and 
$m=3$ links per node.
The traffic was routed using the standard shortest-path strategy, and
the number of data streams, passing a node was counted for each
node. Since by construction at least $m (N-1)$ data channel are open
per node,
we define the load $l_i$, to be the number of data channels divided by
$N$, the number of nodes, at node $i$. In the large $N$ limit this
gives the number of links $m$, if no ``foreign'' connection is going through
the selected node.
One can regard this quantity as ``weighted degree'', with weights
equal to the link loads.

To study the effect of congestion, we assume, that for $N$ nodes a
data stream occupies $1/N$ part of the link capacity. With $N(N-1)$
connections there will be certainly ``overused'' links, where the link
bandwidth capacity constrains the througput of the node. Thus we define the
throughput of a link to be the minimum of the load and the link
capacity, while the throughput of the node to be the sum of the link
troughputs over the links connected to the given node.

Figure~1 shows the tails of the cumulative distribution functions
($1.0-CDF$) for degrees, load and (node) throughput on graphs with
$\alpha$=0, 0.5 and 1.
The distribution of the \emph{degrees}
show the expected tails, exponential ($\alpha=0$, ER graph), 
Weibull ($\alpha=0.5$), and the power-law ($\alpha=1$, BA graph),
respectively.
However, the distribution of load shows interesting deviations from
the one of the nodal degree. 
For $\alpha=0$ the weighted nodal degree (load) distribution can be
approximated much better by a Weibull distribution than an exponential one,
describing the nodal degree distribution. This is a clear indication,
that the distribution of the network traffic even in a simulated
``uniform'' situaton does not follow one-by-one the underlying network
topology, rather developes a heavier tail. For $\alpha=0.5$ the load
remains Weibull, however, the shape parameter (theoretically
$1-\alpha$), changes from the numerical value $0.62$ to $0.32$.
In case of $\alpha=1$ the power-law decay survives, but with changing
CDF tail exponent, descreasing from the numerical value $1.97$ to $1.25$.
We may conclude, that the load distribution has
considerably fatter tails than the underlying nodal distribution.

The per-node \emph{throughput} shows a transition between the nodal
degree and the load distribution. For low throughput values it follows
the load distribution (the bandwidth is enough to hold the traffic),
however, at higher values it approaches the nodal degree distribution, simply
counting the number of links. The transition is governed by the link
capacities, in a real network environment with large available
capacity one expects the througput to follow the load distribution,
however, in an underdesigned network it follows the degree
distribution, as we show in the next section.

\begin{center}
\begin{figure*}
\hspace*{-5mm}
\includegraphics[width=0.35\textwidth]{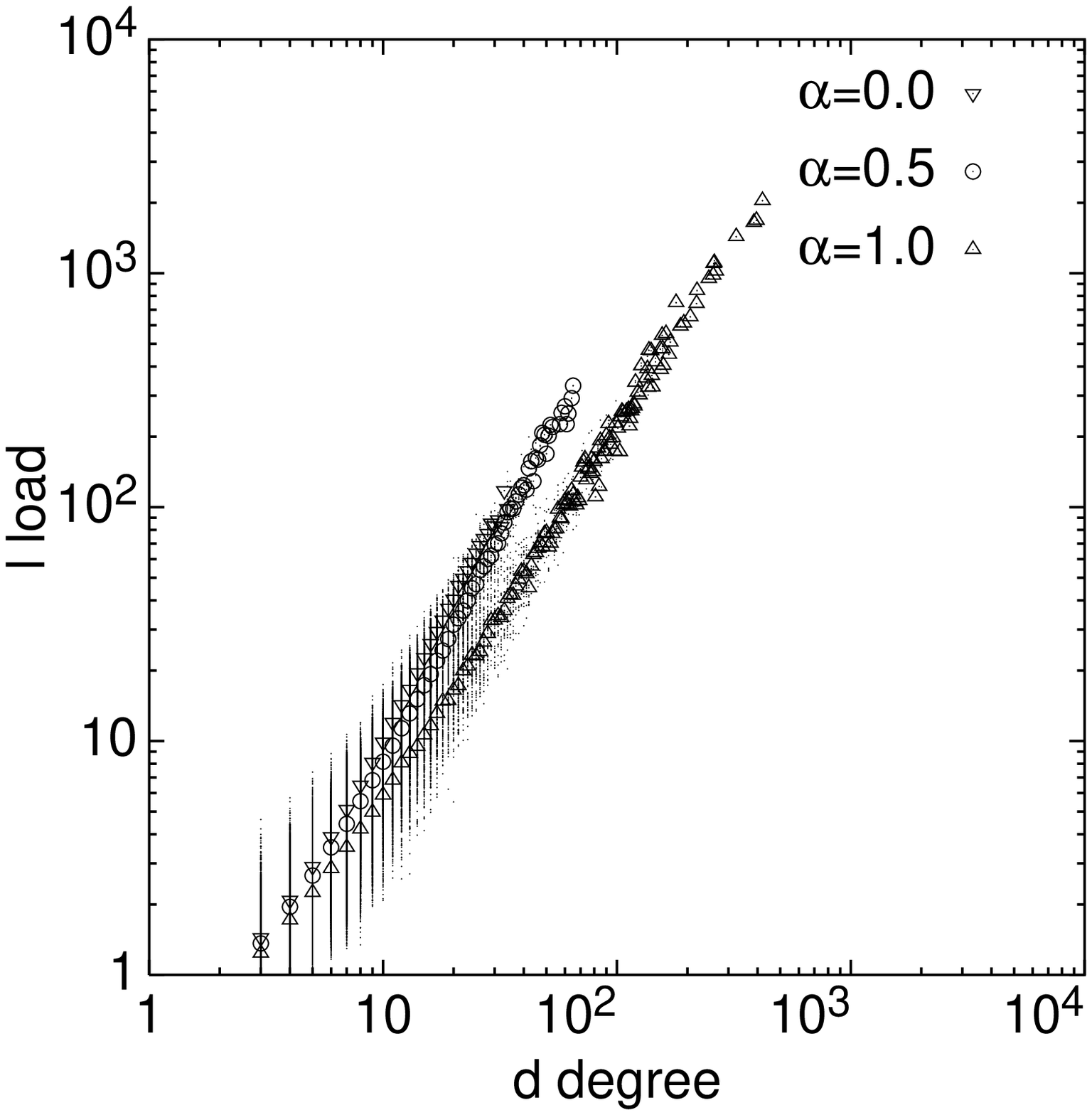}
\hspace*{10mm}
\includegraphics[width=0.345\textwidth]{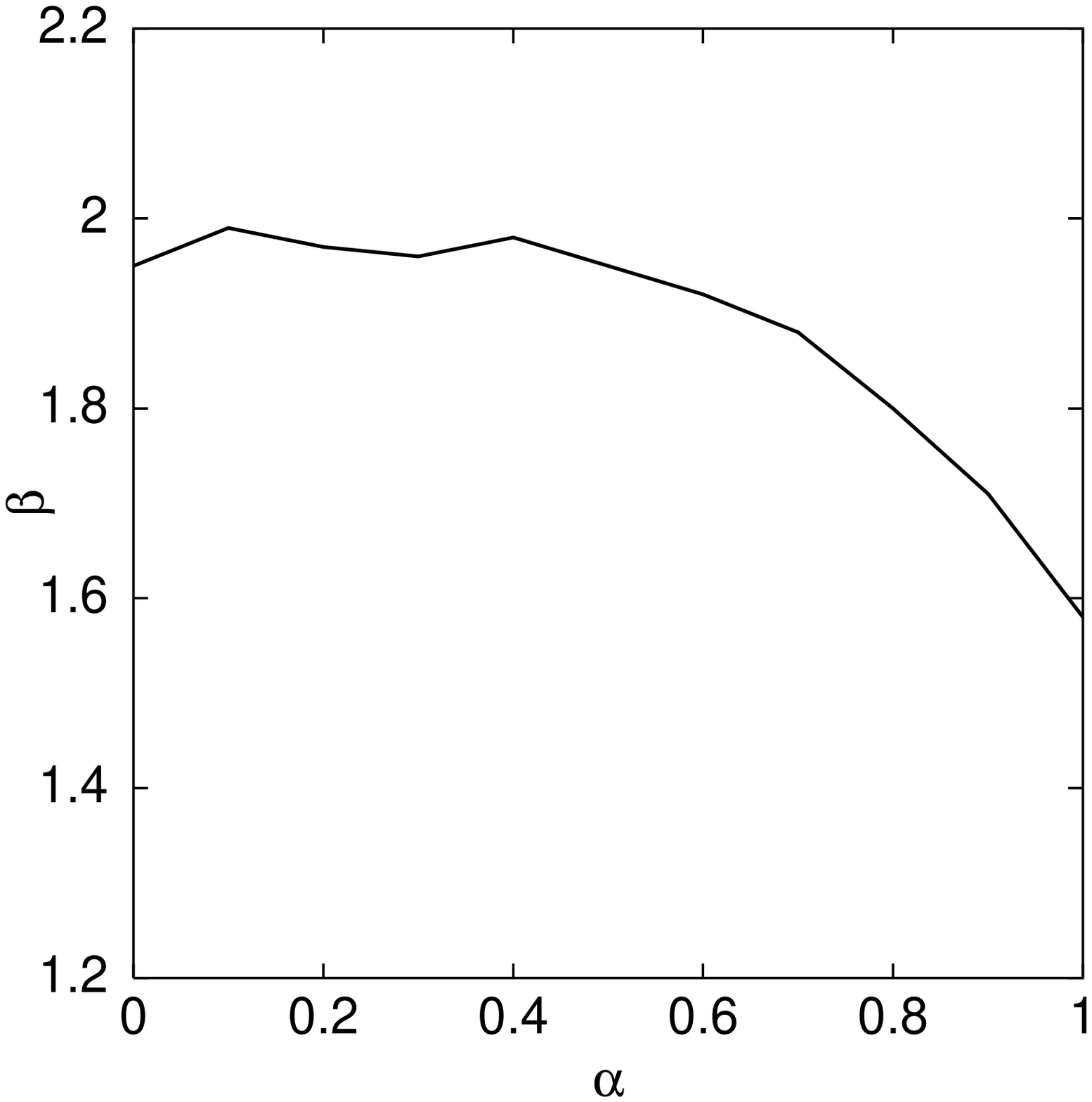}
\caption{The degree-load joint distribution for different network
  geometries (left). Scaling exponent $\beta$, of the load with the
  network geometry parameter $\alpha$ (right).}
\end{figure*}
\end{center}

The change of the distribution from exponential to  Weibull, and the parameter
changes in the Weibull and in the power-law case in the load
distribution suggests to examine the correlation between the degree
and the load. This is shown in Figure~2 (left) on a scatter-plot.
Since it is log-log scaled, the linear clusters indicate a power-law 
correlations, $l_i=d_i^\beta$. The dependence of the scaling exponent
$\beta$, on the network geometry parameter $\alpha$, is shown in
Figure~2 (right). The load is pushed to have fatter tails than the
degree distribution, and the more ``classical'' is the network the
larger the deviation. For the BA geometry the the load distribution
decays $\sim$ 1.6 times faster than the corresponding degree distribution.

This means that not only the distribution of the load is more heavy-tailed
than the distribution of the degrees, but also means that this
dependence is quite strictly a power-law function. It also explains
the deviations from the degree distributions, since a power of
an exponentially distributed ($\alpha=0$) random variable is Weibull,
while in the case of Weibull or power-law degree distributon the
transformation results only in a parameter change for the
distribution. It seems that the traffic pushes the distribution into
heavier tails, from the exponential distribution, and
converging to a power-law through Weibulls. The initial push at
$\alpha=0$ is extremly high, with an exponent 1/2 in the Weibull
distribution. The load distribution is much less sensitive to the
underlying network as the degree distribution.

Since the nodal throughput is bounded from above by the degree, it limits
the throughput on the high-degree nodes, where the links are already
fully utilised. Therefore the throughput-degree joint distribution is
different from the load-degree distribution, the power-law
correlation is only valid for low and medium-degree nodes. But in real
life the congestion at the overloaded nodes also affects the other
parts of the network, since every data flow through these nodes is
jammed. Furthermore, the TCP dynamics is also known to be
chaotic~\cite{pisti} which may change the scalings observed in a
simpler model. To simulate the real life situation, and compare them with the
results of the simplified model, next, we simulate a realistic traffic
on a computer network, too.

\section{Estimation of the network load in a simulated traffic}

In order to compare the theoretical results from the previous section
to a more realistic setup, we 
simulated a TCP/IP network with $N=1024$ nodes, $m=3$ links per node, and a
uniform link bandwidth of $1$ Mb/s. Every node communicated with a randomly
selected target node. To study the effect of congestion we modelled
three scenarios: a low, a medium, and a high traffic one, using
constant bitrate data flows of $16$ Kb/s, $64$ Kb/s, and $256$ Kb/s,
respectively. The simulation was ran using the Berkeley Network
Simulator package~\cite{ns}.
The link throughput was calculated as the number of packets
sent through that link, and the nodal values as the sum of the
throughput of the incident links.

\begin{center}
\begin{figure*}
\hspace*{-5mm}
\includegraphics[width=0.37\textwidth]{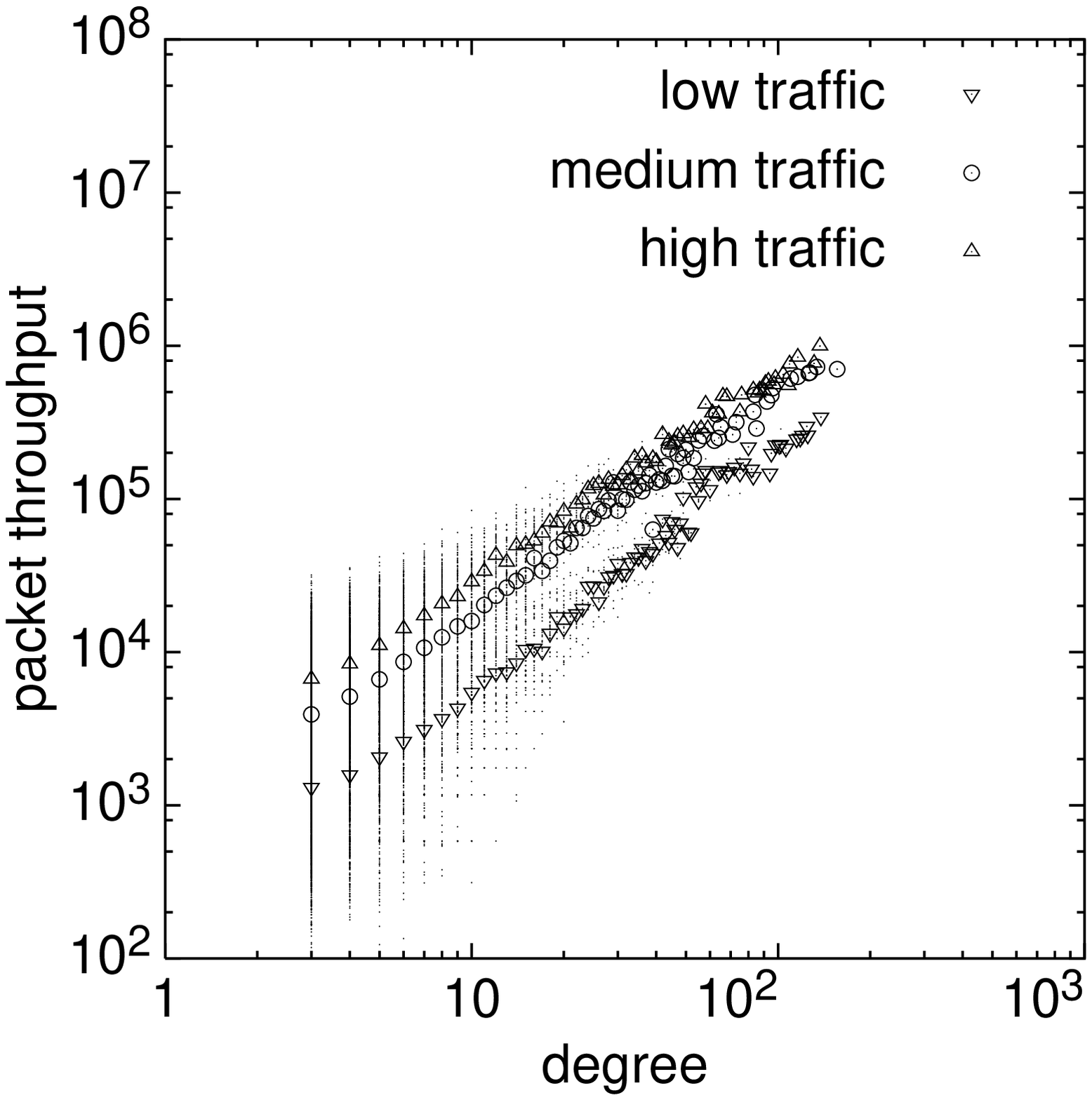}
\hspace*{10mm}
\includegraphics[width=0.35\textwidth]{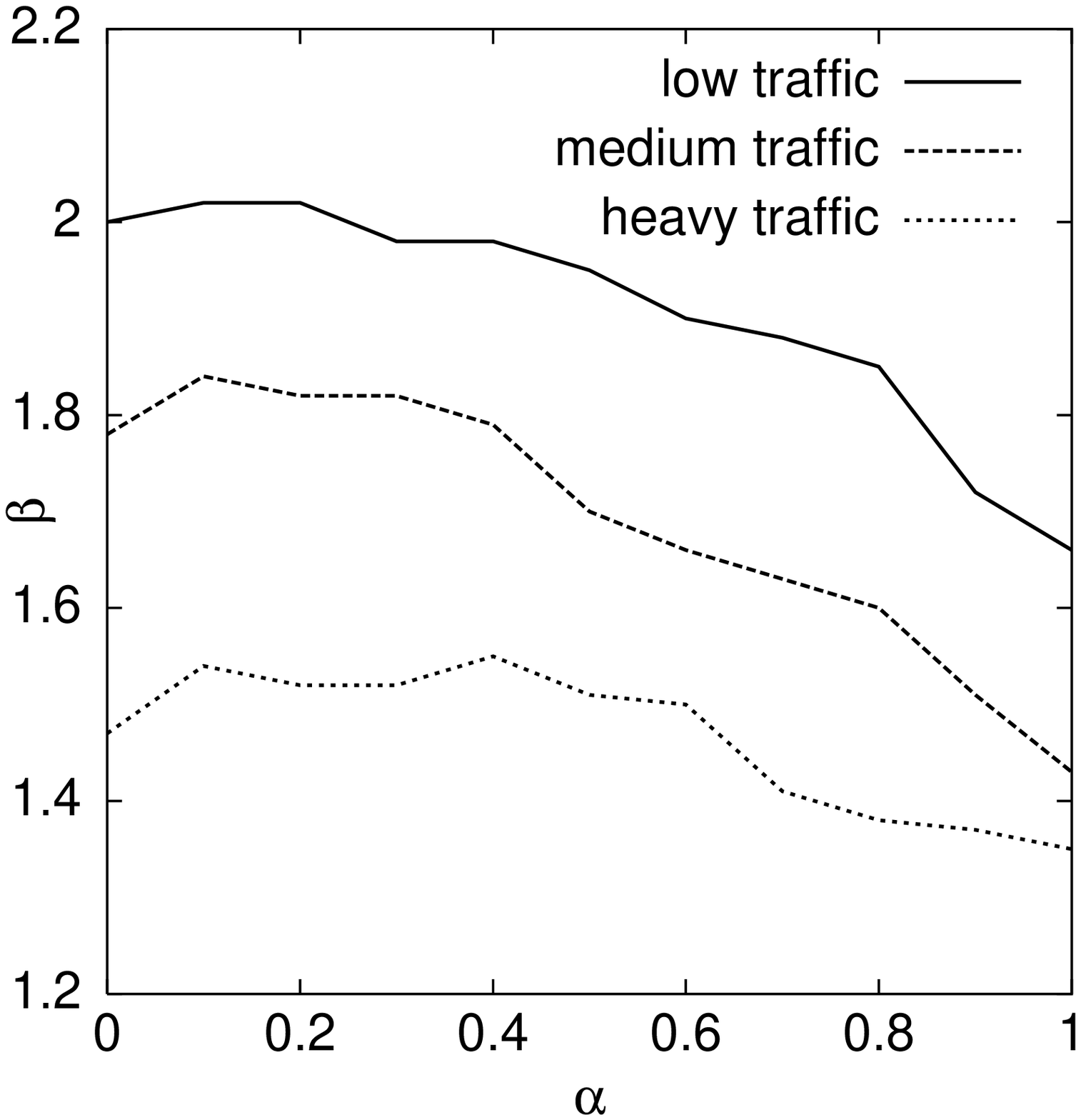}
\caption{Simulated traffic on an $\alpha=1$ geometry network. The
  degree-load joint distribution for different network traffic
  (left). Scaling exponent $\beta$, of the load with the
  network geometry parameter $\alpha$ for low, medium and high traffic(right).}
\end{figure*}
\end{center}

To our surprise, in each scenario the throughput showed a good scaling
through the whole degree range. Since the throughput at the
high-degree nodes is obviously limited, it must mean that the
congestion limits the throughput of the other nodes in such way that
the power-law throughput-degree correlation remains valid. It also
means that the congestion affects each other node proportionally to
its degree.

The correlation exponent $\beta$ was, however, different in the three
simulation. In the low-traffic scenario the measured distribution is
exactly the same as obtained from the numerical simulations of the
load in the previous section. The congestion still has not set up, and
the throughput is identical to the load. As the traffic intensity
grows, exponent $\beta$ decreases, and flattens. It shows that the
heavy traffic is more evenly spreads through the
network, but its dependence on the degrees
remains a power-law function.

\section{Overall performance}

The next quantity we studied is the total number of transferred
packets in the simulated network. Each node transmits constantly TCP
packects to its randomly choosen partner, and if a packet arrives, an
acknowledgement is sent back to the originating node. We counted the
number of packets for which the acknowledgement was received and the
difference betweeen the number of transmitted packets and acknowledged
packets is the loss.

The source of the loss of packets in the network is the congestion:
whether the link capacity cannot handle the amount of traffic, or the
nodes in between cannot cope with the routing of the packages. One
would expect, that in a network with smaller shortest pathes between
two randomly choosen nodes the load on the links and routers is
higher, hence using the same devices it would drop the packets more
often as a network with larger shortest path.

It is also known, that scale free network have smaller shortest pathes
connecting two arbitrary nodes~\cite{abr,cluster}, {\it i.e.} a scale
free network uses less routing devices, however, the load on them is
higher. Indeed, numerical simulation, performed in the previous
section also showed, that with increasing traffic the performance of
the scale free (BA) network downgrades, for example, with a drop rate of 24\%
already at medium traffic, while the classical (ER) network show a
downgrade only of 6\% for the same traffic.

\begin{table}[h]
\begin{tabular}{|l|c|c|c|}
\hline
scenario & $\alpha=0.0$ & $\alpha=0.5$ & $\alpha=1.0$  \\
\hline
low traffic & 0.59 & 0.59 & 0.57 \\
\hline
medium traffic & 2.21 & 2.09 & 1.74 \\
\hline
high traffic & 3.42 & 3.26 & 2.96 \\
\hline
\end{tabular}
\caption{The performance of the networks in millions of packets successfully sent}
\end{table}

\section{Conclusion}

In this paper we showed, 
that the distribution of the network traffic even in a simulated
``uniform'' situaton does not follow one-by-one the underlying
network, rather it developes fatter tails than the nodal degree
distribution of the network. A scaling between the nodal degree
distribution and the load on a node was established, showing a power
like patter, $l\sim d^{\beta(\alpha)}$, where the scaling exponent
$\beta$, is a decreasing function of the network parameter,
$\alpha$. For networks with fatter nodal distribution the exponent is
smaller. As a consequence, attaching a new connection to the node
requires less upgrade in the bandwidth for a scale free network to
keep the performance, as for a classical (ER) one.

The above theoretical result was confirmed by simulation, the per-node
throughput is still {\em scaling} with nodal degree. The scaling
dependes on the amount of the traffic, for a completely congested
situation the throughput distribution by definition agrees with the
nodal degree distribution, hence the scaling exponent is 1, however,
for partly congested or congestion free networks this scaling
approaches the load distribution with scaling exponents in the range
1.4 (partially congested BA network) to 2 (congestion free ER). A
scale free network requires less resource upgrade when a new node is
added. 

The overall performance of a scale free network is decreasing rapidly
with the traffic, where the classical network still has almost no
loss. It is due to the feature, that the ER network uses more routers
along the shortest connection, and hence the traffic is distributed
more evenly.

\end{document}